\documentclass[11pt,a4paper]{article}
\usepackage{amsmath,amssymb}
\usepackage{bm}
\usepackage{graphicx}
\usepackage{enumerate}
\usepackage{ascmac}
\usepackage{diagbox}
\usepackage{multirow}
\usepackage{threeparttable}
\usepackage{tikz} 
\usepackage{pgfplots} 
\usepackage{float}
\usepackage{authblk}
\usepackage{caption}
\usepackage{apacite}
\usepackage{geometry}
\newcommand{\argmin}{\mathop{\rm argmin}\limits}
\title{Positive-Unlabelled Survival Data Analysis}
\author[1]{Tomoki Toyabe}
\author[1]{Yasuhiro Hasegawa}
\author[2,3]{Takahiro Hoshino}

\affil[1]{Graduate School of Economics, Keio University, Tokyo, Japan}
\affil[2]{Faculty of Economics, Keio University, Tokyo, Japan}
\affil[3]{RIKEN AIP, Tokyo, Japan}

\date{\today}
\begin{document}
\maketitle
\begin{abstract}
In this paper, we consider a novel framework of positive-unlabeled data in which as positive data survival times are observed for subjects who have events during the observation time as positive data and as unlabeled data censoring times are observed but whether the event occurs or not are unknown for some subjects. We consider two cases: (1) when censoring time is observed in positive data, and (2) when it is not observed. For both cases, we developed parametric models, nonparametric models, and machine learning models and the estimation strategies for these models. Simulation studies show that under this data setup, traditional survival analysis may yield severely biased results, while the proposed estimation method can provide valid results.
\end{abstract}

\section{Introduction}
Predicting mortality in acute and high-risk patients and discovering the variables that cause the disease are important issues in the medical field. In addition, in the field of marketing, it is a top priority to find out why users of a company's services are leaving and take appropriate measures to keep them there. In this paper, we will refer to these two types of targets as "users". Survival analysis is an effective method for solving these issues. Survival analysis allows us to examine factors associated with survival time, and the model we have created can be used to predict the expected survival time of surviving patients and users of services on an ongoing basis. In the past, previous research has used survival time analysis in the medical field to investigate the impact of obesity on life expectancy in the United States in the 21st century showed by \shortciteA{olshansky2005potential} and factors associated with coronary syndrome, a type of heart disease, on a large scale proposed by \shortciteA{amsterdam20142014}. In addition in the field of marketing, \shortciteA{reinartz2003impact} associated factors that extend the duration of service with the amount of purchase, and \shortciteA{dadzie2018trust} used survival analysis to conclude that cognitive and emotional factors are involved in the duration of interpersonal trust between buyers and sellers.
\par
In the field of statistical method development, censoring occurs when the observed data falls below a certain threshold. Censored data are not counted as observational data and this can cause a large bias in the estimates. Various methods have been developed to deal with this problem as a regression method, including the Tobit model suggested by \shortciteA{tobin1958estimation}. A similar concept of censoring exists in theoretical aspects of survival time analysis. The concept of right-side censoring is the most major censoring problem in survival time analysis. Right-sided censoring is a problem in which the time $t_i$ to true death is missing when the observed duration $t_i$ is longer than the censoring time $c_i$, which indicates the time to observe the data (i.e., when the user is still alive). Many methods of survival analysis have been developed in the past as models that take the above events into account. \shortciteA{lee2003statistical} and \shortciteA{yamaguchi1991event}  proposed models assuming  Weibull, exponential and normal distributions for the hazard function $f(x)$ as a parametric method for survival analysis. \shortciteA{ibrahim2014b} proposed a bayesian model to estimate these distributions. \shortciteA{cox1972regression} suggested the Cox proportional hazards model, which calculates the hazard ratios in semi-parametric models, is well known, and \shortciteA{su2016sparse} suggested the sparse Cox proportional hazards model, which is based on the Cox proportional hazards model and corresponds to the sparse positive of the data, and the magnitude of covariance, has been proposed as Lasso Cox proportional hazards suggested by \shortciteA{zhang2007adaptive}, which corrects and calculates the true parameters. Finally, in non-parametric methods, \shortciteA{efron1988logistic} suggested Kaplan-Meier models, and \shortciteA{muller2004nonparametric} proposed bayesian non-parametric models. Methods using machine learning methods belonging to non-parametric models have also been developed, including a model for survival analysis using Neural Network suggested by \shortciteA{liestbl1994survival}, a model for survival analysis using decision trees suggested by \shortciteA{segal1997features}, a model for survival analysis using random forests, and a method for models of survival analysis that have been developed in the past suggested by \shortciteA{ishwaran2008random}, models of survival analysis that incorporate random forests corresponding to imbalance data suggested by \shortciteA{afrin2018balanced}, and models of survival analysis with multitasking type processing suggested by \shortciteA{li2016multi}.\par
However, existing models for survival analysis are not able to deal with the PU structure where only a portion of the observed censoring labels are known for positive cases, and the rest are a mixture of positive and negative cases. A model for survival analysis for the positive unlabeled data has not been developed. In the past, \shortciteA{lancaster1996case} proposed the assumption $p(x|s_i=1)=p(x|y_i=1)$,$p(x|s_i=0)=p(x)$ for the data structure, using a variable $s_i$ to indicate whether the data are positive or Unlabeled, when dealing with the binary classification problem of variable ${y_i}\in\{0,1\}$, which represents the assignment of groups of $i$ to each user.

On the other hand, \shortciteA{elkan2008learning} proposed a framework for PU learning using two assumptions, $p(s_i=1|x,y_i=0)=0$ and $p(s_i=1|x,y_i=1)=p(s_i=1|y_i=1)$. The latter assumption, however, used a strong assumption that the probability of becoming D in C is irrelevant to $x$, and thus did not fit the real data and survival time. However, the latter of the two assumptions proposed by Elkan et al. used the strong assumption that the probability of becoming $s_i=1$ in $y_i=1$ is irrelevant to $x$, an assumption that does not fit the real data and is not applicable to the data handled by survival analysis.

In this paper, we extend the two assumptions proposed by Lancaster and Imbens (1996) to survival time analysis and propose a model for survival time analysis that takes into account situations where the information is inconclusive. Chapter 2 presents the assumptions of the setup and data structure assumed in this thesis. We derive the likelihood function in Chapter 3, present the simulation setup and results in Chapter 4, and conclude this thesis with the conclusion in Chapter 5.

\section{Model}
\subsection{Setup}
In this section, we define the data structure when the data assumed by this paper are not definitive. Consider a survival time analysis with censoring, where $t$ is the survival time and $c$ is the censoring time. Let $y$ be the indicator of censoring as follows.
\begin{equation}
\label{indicator}
y=1(t<c)
\end{equation}

Most survival analyses have assumed that $t>c$ is always true in the case of censoring. \shortciteA{miller2011survival} In this study, however, we assume that $t$ is simply unobservable when censoring occurs, regardless of the size of $t$ relative to $c$. This is a common situation in the real world, but it has not been established as a model for survival analysis. The fact that the size of $t$ relative to $c$ is unknown when censoring occurs means that ''the data are Unlabeled''. To help understanding, we assume a situation in which we make a sale to each customer from a customer list. In this case, the notation $i$ is used to identify each customer. The observed survival time, censoring time, and censoring indicator can be expressed as $\{t_i,c_i,y_i\}$. In a sales situation, if a sale can be made, $t_i$ is recorded at that point and removed from the list as $y_i=1$. At this time, this means that a customer $i$ who did not become $y_i=1$ by the sales period $c_i$ was purchasing from a competitor in the same industry for which the company is unable to obtain data, etc., is not necessarily $t_i > c_i$.\newline
Using the variable $s_i$, $s_i = 1$ for the user $i$ who is known to have purchased from the company and $s_i = 0$ for other customers. If $s_i = 1$, then $y_i = 1$, and if $s_i = 0$, then $y_i = 0$ or $1$, and the framework known as Presence-Only Data and PU Learning can also be used in survival analysis.

In this paper, we focus on both cases where $c_i$ is observed when $s_i=1$ and when $c_i$ is not observed. We set $x_i$ as a variable that represents the unique information of user $i$. When $c_i$ is observed, ($t_i,c_i,x_i$) is observed when $s_i=1$, and when $c_i$ is not observed, ($t_i,x_i$) is only observed when $s_i=1$. In both cases, only ($c_i,x_i$) is assumed to be observed when $s_i=0$.When $s_i=1$, $c_i$ is observed when, for example, the time to discontinue sales activities is set in advance when the sales are made to user $i$.This is famous as a sales technique for goods with small sales volume and high maintenance cost, such as cars. On the other hand, when $s_i=1$ and $c_i$ is not observed, the opposite is true in the case where no sales termination time is set in advance. This is one of the most common situations in which a company has a sales policy for mobile phone contracts with a large number of sales.

Based on the above setup, the expected dataset for each case is as follows.\par
\begin{table}[htb]
\begin{center}
\begin{tabular}{c}

\begin{minipage}{0.45\hsize}
\begin{center}
\caption{The case where $c_i$ is observed when $s_i=1$}
\begin{tabular}{|c|c|c|c|c|} \hline
ID & $t_i$ & $y_i$ & $c_i$ & $s_i$ \\ \hline
1 & 80 & 1 & 90 & 1 \\
2 & 90 & 1 & 100 & 1 \\
3 & ? & ? & 70 & 0 \\
4 & 70 & 1 & 90 & 1 \\
5 & ? & ? & 60 & 0 \\ \hline
\end{tabular}
\label{tab:price1}
\end{center}
\end{minipage}

\begin{minipage}{0.45\hsize}
\begin{center}
\caption{The case where $c_i$ is unobserved when $s_i=1$}
\begin{tabular}{|c|c|c|c|c|} \hline
ID & $t_i$ & $y_i$ & $c_i$ & $s_i$ \\ \hline
1 & 80 & 1 & ? & 1 \\
2 & 90 & 1 & ? & 1 \\
3 & ? & ? & 70 & 0 \\
4 & 70 & 1 & ? & 1 \\
5 & ? & ? & 60 & 0 \\ \hline
\end{tabular}
\label{tab:price2}
\end{center}
\end{minipage}
\end{tabular}
\end{center}
\end{table}
In this section, we have described the data structure for the case where $c_i$ is observed and the case where $c_i$ is not observed when $s_i=1$. In the following section, we present the assumptions needed to estimate the model and the settings for censoring without information.
\subsection{Setting}
In this section, we present the following two assumptions for the estimation of the model proposed by this paper.
\subsubsection{Two assumptions about data structure}
At first, we make the following two assumptions about $t_{i},c_{i}$ to correspond to the $s_i$ assumed in the data structure proposed in this paper.
\begin{enumerate}
\item In the case where $s_i=1$ is labeled, $t_i,c_i$ is assumed to be randomly extracted from the $t_i,c_i$ layer conditioned by $t_i<c_i$.
\item In the case where $s_i=0$ is labeled, $t_i,c_i$ is assumed to be randomly extracted from the whole set.
\end{enumerate}
The above assumptions about $t_{i},c_{i}$ can be expressed as follows.
\begin{align}
\label{assumption1}
p(t_i, c_i|x_i,s_i=1)&=p(t_i, c_i|x_i,y_i=1)=p(t_i, c_i|x_i, t_i<c_i)\\
\label{assumption2}
p(t_i, c_i|x_i,s_i=0)&=p(t_i, c_i|x_i)
\end{align}
This assumption holds for a single variable such as $t_i,c_i$, respectively.\par
\subsubsection{Non-informative censoring}
In addition to the settings shown above, this proposed model for survival analysis assumes non-informative censoring. Non-informative censoring is a setting that indicates that each user's survival time $t_i$ and censoring time $c_i$ are not correlated with each other. It is a setting that allows us to separate the joint distribution of $t_i,c_i$ as shown in the following equation.
\begin{equation}
p(t_i,c_i|x_i)=p_t(t_i|x_i)p_c(c_i|x_i)
\end{equation}
The following noteworthy points are noted. In the conventional model of survival time analysis, the distribution of $t$ was not affected by the distribution of $c$ when estimating the distribution of $t$ under the assumption of non-informative censored data. However, in the data structure assumed in this paper, it is possible to deal efficiently with positive unlabeled data by assuming that the distribution of $t$ is affected by the distribution of $c$ when the distribution of $t$ is identified.\par

As a corollary, under the above two assumptions, $s_i$ has a structure that depends not only on $x_i$ but also on $t_i,c_i$. From the above discussion, we cannot use the method of survival time analysis with a weighted average of $p(s_i|x_i)$, which is assumed when $s_i$ depends only on $x_i$. In Chapter 3, we show the likelihood function of the model by using the two assumptions for the data structure presented in this section and the setting of non-informative censoring.

\section{Likelihood}
In this section, we propose the likelihood of the model in the previous section.
\subsection{The case where c is observed}
First, consider the case where censoring time $c$ is observed when $s=1$. At this time, the likelihood is expressed as 
\begin{equation}
L(\theta_t, \theta_c|.)\propto \prod_{i=1}^n p(t_i, c_i| s_i=1, x_i, \theta_t, \theta_c)^{s_i}p_c(c_i | s_i = 0, x_i, \theta_c)^{1-s_i},
\end{equation}
where $p(\cdot)$ is the probability density function and $\theta_t, \theta_c$ are the parameters for the probability density function of $t$ and $c$, respectively. Using the hypothetical formula 
(\ref{assumption1}) and (\ref{assumption2}), the likelihood can be rewritten as 
\begin{equation}
L(\theta_t, \theta_c|.)\propto \prod_{i=1}^n p(t_i, c_i | y_i=1, x_i, \theta_t, \theta_c)^{s_i}p_c(c_i | x_i, \theta_c)^{1-s_i}.
\end{equation}
In this setting, since the joint distribution $p(t_i, c_i | y_i=1, x_i, \theta_t, \theta_c)$ is considered to be a truncated distribution at $t_i<c_i$, the likelihood can be rewritten further as follows.
\begin{align}
L(\theta_t, \theta_c|.)\propto&\prod_{i=1}^n p(t_i, c_i | y_i=1, x_i, \theta_t, \theta_c)^{s_i}p_c(c_i | x_i, \theta_c)^{1-s_i} \nonumber \\
=& \prod_{i=1}^n \left\{\frac{p(t_i,c_i|x_i, \theta_t, \theta_c)}{p(y_i=1|x_i, \theta_t, \theta_c)}\right\}^{s_i}p_c(c_i | x_i, \theta_c)^{1-s_i} \nonumber \\
=& \prod_{i=1}^n \left\{\frac{p_t(t_i|x_i, \theta_t)p_c(c_i|x_i, \theta_c)}{1-\int_0^{\infty} p_t(t_i|x_i,\theta_t)F_c(t_i|x_i,\theta_c)dt_i
}\right\}^{s_i}p_c(c_i | x_i, \theta_c)^{1-s_i} \nonumber \\
=& \prod_{i=1}^n \left\{\frac{p_t(t_i|x_i, \theta_t)}{1-\int_0^{\infty} p_t(t_i|x_i,\theta_t)F_c(t_i|x_i,\theta_c)dt_i}\right\}^{s_i}p_c(c_i|x_i,\theta_c)
\end{align}
Therefore, the likelihood functions of $\theta_t$ and $\theta_c$ can be written as
\begin{align}
L(\theta_t|.) \propto \prod^n_{i=1} \left \{\frac{p_t(t_i|x_i, \theta_t)}{1-\int_0^{\infty} p_t(t_i|x_i,\theta_t)F_c(t_i|x_i,\theta_c)dt_i}\right \}^{s_i},\\
L(\theta_c|.) \propto \prod^n_{i=1} \left \{\frac{1}{1-\int_0^{\infty} p_t(t_i|x_i,\theta_t)F_c(t_i|x_i,\theta_c)dt_i}\right \}^{s_i} p_c(c_i|x_i, \theta_c).
\end{align}
Considering the probability density function of $t_i$ and $c_i$ as the exponential distribution with the parameters $\lambda_{it}$ and $\lambda_{ic}$, respectively, the regression model whose parameters depend on $x$ is written as
\begin{align}
\label{survival}
p_t(t_i | \bm{x}_i, \bm{\theta}_t) &= \lambda_{it} \exp(-\lambda_{it} t_i), \quad \lambda_{it} = \exp(\bm{x}_i' \bm{\theta}_t),\\
\label{censoring}
p_c(c_i | \bm{x}_i, \bm{\theta}_c) &= \lambda_{ic} \exp(-\lambda_{ic} c_i), \quad \lambda_{ic} = \exp(\bm{x}_i' \bm{\theta}_c),
\end{align}
where $\bm{\theta}_t$, $\bm{\theta}_c$ and $\bm{x}_i$ are $p \times 1$ vectors.
The likelihoods can be written as 
\begin{align}
L(\theta_t|.) \propto \prod^n_{i=1} \{(\lambda_{it}+\lambda_{ic}) \exp(-\lambda_{it} t_i)\}^{s_i}, \nonumber \\
L(\theta_c|.) \propto \prod^n_{i=1} (\lambda_{it}+\lambda_{ic})^{s_i}\lambda_{ic} \exp(-\lambda_{ic} c_i).
\end{align}

\subsection{The case where c is unobserved}
Next, consider the case where $c$ is not observed when $s=1$. In the same way, the likelihood can be written as 
\begin{align}
L(\theta_t, \theta_c|.)\propto& \prod_{i=1}^n p_t(t_i| s_i=1, x_i, \theta_t,\theta_c)^{s_i}p_c(c_i | s_i = 0, x_i, \theta_c)^{1-s_i} \nonumber \\
=& \prod_{i=1}^n p_t(t_i| y_i=1, x_i, \theta_t,\theta_c)^{s_i}p_c(c_i | x_i, \theta_c)^{1-s_i} \nonumber \\
=& \prod_{i=1}^n \left\{\frac{p_t(t_i|x_i,\theta_t) \{1-F_c(t_i|x_i,\theta_c)\}}{p(y_i=1|x_i, \theta_t,\theta_c)}\right\}^{s_i}p_c( c_i | x_i, \theta_c)^{1-s_i} \nonumber \\
=& \prod_{i=1}^n \left\{\frac{p_t(t_i|x_i,\theta_t) \{1-F_c(t_i|x_i,\theta_c)\}}{1-\int_0^{\infty} p_t(t_i|x_i,\theta_t)F_c(t_i|x_i,\theta_c)dt_i}\right\}^{s_i}p_c(c_i|x_i, \theta_c)^{1-s_i}.
\end{align}
Similarly, considering the probability density function of $t,c$ as the exponential distribution of the parameters $t_i,c_i$ and $\lambda_{it},\lambda_{ic}$, respectively, the likelihoods can be written as 
\begin{align}
\label{t_likelihood}
L(\theta_t|.) &\propto \prod^n_{i=1} \left\{(\lambda_{it}+\lambda_{ic})\exp(-\lambda_{it} t_i)\right\}^{s_i},\\
\label{c_likelihood}
L(\theta_c|.) &\propto \prod^n_{i=1} \{(\lambda_{it}+\lambda_{ic})\exp(-\lambda_{ic} t_i)\}^{s_i}\{\lambda_{ic} \exp(-\lambda_{ic} c_i)\}^{1-s_i}.
\end{align}
In the chapter 5, we simulate both cases where $c_i$ is observed and where $c_i$ is unobserved using parametric simulations to show that the proposed model is appropriate.

\section{Additional Settings}
\subsection{Gamma and Weibull distributions}
Considering the probability density function of $t_i$ as the Gamma distribution with the parameters $\alpha_t, \lambda_{it}$ and that of censoring time as the exponential distribution with the parameter $\lambda_{ic}$, the regression model whose parameters depend on $x$ is written as
\begin{align}
p_t(t_i | \bm{x}_i, \alpha_t, \bm{\theta}_t) &= \frac{1}{\Gamma(\alpha_t)}t_i^{\alpha_t-1}\lambda_{it}^{\alpha_t } \exp( -\lambda_{it} t_i), \quad \lambda_{it} = \exp(\bm{x}_i' \bm{\theta}_t), \\
p_c(c_i | \bm{x}_i, \bm{\theta}_c) &= \lambda_{ic} \exp(-\lambda_{ic} c_i), \quad \lambda_{ic} = \exp(\bm{x}_i' \bm{\theta}_c),\\
F_c(c_i | \bm{x}_i, \alpha_c, \bm{\theta}_c)& = 1 - \exp(-\lambda_{ic}c_i),
\end{align}
where $\bm{\theta}_t$, $\bm{\theta}_c$ and $\bm{x}_i$ are $p \times 1$ vectors. In this condition, we can solve $p(y_i=1|\bm{x}_i, \bm{\theta}_t, \bm{\theta}_c)$ analytically as
\begin{align}
&1-\int_0^{\infty} p_t(t_i | \bm{x}_i, \alpha_t, \bm{\theta}_t)F_c(t_i|\bm{x}_i, \alpha_c, \bm{\theta}_c)dt \nonumber \\
= &1-\int_0^{\infty} p_t(t_i | \bm{x}_i, \alpha_t, \bm{\theta}_t)(1 - \exp(-\lambda_{ic}t_i))dt \nonumber \\
= &1-\frac{1}{\Gamma(\alpha_t)}\int_0^{\infty} t_i^{\alpha_t-1}\lambda_{it}^{\alpha_t } \exp( -\lambda_{it}t_i)(1 - \exp(-\lambda_{ic}t_i))dt \nonumber \\
=&1-\frac{1}{\Gamma(\alpha_t)}\left[\left(\frac{\lambda_{it}}{\lambda_{it}+\lambda_{ic}}\right)^{\alpha_t}\Gamma(\alpha_t,(\lambda_{it}+\lambda_{ic})t)-\Gamma(\alpha_t,\lambda_{it}t)\right]_0^{\infty} \nonumber \\
=&1-\frac{1}{\Gamma(\alpha_t)}\left\{\Gamma(\alpha_t)-\left(\frac{\lambda_{it}}{\lambda_{it}+\lambda_{ic}}\right)^{\alpha_t}\Gamma(\alpha_t)\right\} \nonumber \\
=&\left(\frac{\lambda_{it}}{\lambda_{it}+\lambda_{ic}}\right)^{\alpha_t}
\end{align}
where $\Gamma(a,b)$ is the incomplete gamma function (second-type) and $\Gamma(a)$ is the gamma function.
If the distribution of $t$ is  a Weibull distribution or the distribution of $c$ is also a gamma distribution or a Weibull distribution, the probability cannot be calculated analytically. This is discussed in Appendix 2. 

\subsection{Extension to Nonparametric bayesian model}
In order to express more general distribution for survival time $t$, we can consider the probability density function of $t$ as the Dirichlet process Gamma mixture model. Let the Dirichlet process priors $G \sim DP(\alpha, G_0)$, G can be expressed as 

\begin{equation}
G = \sum_{k=1}^{\infty} \pi_k \delta_{\phi_k}, \quad \phi_k \sim G_0,
\end{equation}

where $k$ means the number of mixture components, $\phi_k$ is the distribution parameter of the $k$-th mixture component, $\delta_{\phi_k}$ is the dirac delta measure supported at $\phi_k$, $G_0$ is the baseline measure and $\pi_k$ denotes a weight for the $k$-th mixture component defined as

\begin{equation}
\pi_1 = V_1, \quad \pi_k = V_k \prod_{i=1}^{k-1}(1-V_i), \quad V_k \sim Beta(1, \alpha).
\end{equation}

In this setting, the probability density function of $t$ can be written as 
\begin{equation}
p_t(t_i | x_i, \Phi) = \sum_{k=1}^{\infty} \pi_k p_t(t_i | x_i, \phi_k).
\end{equation}

Considering the $k$-th component of the probability density function of $t_i$ as the Gamma distribution with the parameters $\alpha_kt, \lambda_{kit}$ and that of censoring time as the exponential distribution with the parameter $\lambda_{ic}$, the regression model whose parameters depend on $x$ is written as
\begin{align}
p_t(t_i | \bm{x}_i, \alpha_t, \bm{\theta}_t) &= \sum_{k=1}^{\infty} \pi_k \frac{1}{\Gamma(\alpha_{kt})}t_i^{\alpha_{kt}-1}\lambda_{kit}^{\alpha_{kt} } \exp( -\lambda_{kit} t_i), \quad \lambda_{kit} = \exp(\bm{x}_i' \bm{\theta}_{kt}), \\
p_c(c_i | \bm{x}_i, \bm{\theta}_c) &= \lambda_{ic} \exp(-\lambda_{ic} c_i), \quad \lambda_{ic} = \exp(\bm{x}_i' \bm{\theta}_c),\\
F_c(c_i | \bm{x}_i, \alpha_c, \bm{\theta}_c)& = 1 - \exp(-\lambda_{ic}c_i).
\end{align}

Similarly, we can solve $p(y_i=1|x_i, \theta_t, \theta_c)$ analytically as
\begin{align*}
&1-\int_0^{\infty} \left(\sum_{k=1}^\infty \pi_k p_t(t_i | \bm{x}_i, \alpha_{kt}, \bm{\theta}_{kt})\right)F_c(t_i|\bm{x}_i, \alpha_c, \bm{\theta}_c)dt\\
= &1-\int_0^{\infty} \left(\sum_{k=1}^\infty \pi_k p_t(t_i | \bm{x}_i, \alpha_{kt}, \bm{\theta}_{kt})\right)(1 - \exp(-\lambda_{ic}t_i))dt\\
= &1-\int_0^{\infty} \sum_{k=1}^\infty \pi_k \frac{1}{\Gamma(\alpha_{kt})} t_i^{\alpha_{kt}-1}\lambda_{kit}^{\alpha_{kt} } \exp( -\lambda_{kit}t_i)(1 - \exp(-\lambda_{ic}t_i))dt\\
=&1-\sum_{k=1}^\infty \pi_k\frac{1}{\Gamma(\alpha_{kt})}\left[\left(\frac{\lambda_{kit}}{\lambda_{kit}+\lambda_{ic}}\right)^{\alpha_{kt}}\Gamma(\alpha_{kt},(\lambda_{kit}+\lambda_{ic})t)-\Gamma(\alpha_{kt},\lambda_{kit}t)\right]_0^{\infty}\\
=&1-\sum_{k=1}^\infty \pi_k\frac{1}{\Gamma(\alpha_{kt})}\left\{\Gamma(\alpha_{kt})-\left(\frac{\lambda_{kit}}{\lambda_{kit}+\lambda_{ic}}\right)^{\alpha_{kt}}\Gamma(\alpha_{kt})\right\} \\
=&\sum_{k=1}^\infty \pi_k \left(\frac{\lambda_{kit}}{\lambda_{kit}+\lambda_{ic}}\right)^{\alpha_{kt}} \quad \left(\because \sum_{k=1}^\infty \pi_k = 1 \right),
\end{align*}
where $\bm{\theta}_{kt}$, $\bm{\theta}_c$ and $\bm{x}_i$ are $p \times 1$ vectors. According to \shortciteA{ishwaran2001gibbs}, we assume the number of the mixture components as the sufficient large finite number K in the MCMC scheme and therefore we can express the likelihood in the closed form.

\subsection{Extension to Machine Learning}
In this section, we explain how the proposed framework can be extended to machine learning methods by defining a loss function. Loss functions are proposed for the cases where the observed time is continuous and discrete time, respectively. The Cox proportional hazards model is considered for continuous time and a discrete survival logit model is considered for discrete time.

\subsubsection{The case where the observation time is continuous}
The Cox proportional hazards model is considered as a model for continuous time survival analysis and a semiparametric model of survival analysis in which the hazard function is defined below by assuming proportional hazard suggested by \shortciteA{cox1972regression}.
\begin{align}
h(t_i|x_i)=h_0(t_i)g(x_i),\ g(x_i)=\theta_t{x_i},
\end{align}
where $x_i$ is a covariate vector and $\theta_t$ is a parameter vector. This model consists of a non-parametric function $h_0(t_i)$, and a parametric function $g(x_i)$, which can be estimated using the following partial likelihood
\begin{align}
L_{cox}=\prod_{i=1}\left(\frac{\exp[g(x_i)]}{\sum_{j\in{R_i}}\exp[g(x_j)]} \right)^{y_i},
\end{align}
where $R_i$ is the set of users whose observations continue without censoring at time $t_i$. Kvamme et al. (2019) extended the Cox proportional hazards model to neural networks by proposing a loss function with a partial likelihood $L_{cox}$ of negative logarithmic partial likelihood when $g(x_i)$ is a nonlinear function. \shortciteA{kvamme2019time} To define a loss function $loss_{cox}$ for our model, we can define the following loss function as
\begin{equation}
\label{losscox}
loss_{(cox)}=\frac{1}{n_{(s_i=1)}}\sum_{i=1}^n \left\{s_i\log\left(\sum_{j\in{R_i}}\exp[g(x_j)-g(x_i)]\right)\times\frac{1}{p(y_i=1|x_i)}\right\},
\end{equation}
\begin{equation}
\label{ycox}
{p(y_i=1|x_i)}=1-\int_0^{\infty} p_t(t_i | x_i, \theta_t)F_c(t_i|x_i, \theta_c)dt
\end{equation}
where $g(x_i)$ is a nonlinear function and $n_{(s_i=1)}$ represents the sample size of the data with $s_i=1$. We assume that the distribution of censoring times is known to further estimate $\theta_t$.The likelihood of our model shows that the model can be estimatied using only the data of $s_i=1$ and that the loss function using only the data of $s_i=1$ and weighted by the inverse of $p(y_i=1|x_i)$ can be regarded as the overall loss function. The method for estimating $\theta_t$ using this loss function is presented in the section \ref{mlest}.
\subsubsection{The case where the observation time is discrete}
The discrete survival logit model proposed by \shortciteA{singer1993s} is targeted as a model for discrete time survival analysis.
We define a variable $\tau$ that represents a discrete time. And if the maximum time of the observation is $J$, then the time of $\tau$ satisfies $\tau\in\mathcal{N}$ and $1\leqq \tau \leqq J$. At this time, the hazard function of the discrete survival logit model for user $i$ at time $\tau$ can be defined as
\begin{align}
h(\tau|x_i)=\frac{1}{1+\exp(-(\bm{\alpha}_t'\bm{D}+\bm{\beta}_t'\bm{x_i}))},
\end{align}
where $\bm{\alpha}_t$ represents the sequence of intercept terms of the hazard function for each time and $\bm{D}$ is a dummy variable that chooses the intercept corresponding to time $\tau$ from among the $\bm{\alpha}_t$, $x_i$ is a covariate vector and $\beta_t$ is a parameter vector.\\
Defining the loss function $loss_{logit}$ to find $\theta_t=(\bm{\alpha}_t,\bm{\beta}_t)$ corresponding to the framework proposed in this paper as well as the case of continuous time in the previous section is as
\begin{equation}
\label{losslogit}
loss_{(logit)}=\frac{1}{n_{(s_i=1)}}\sum_{i=1}^n \left\{s_i\sum_{k=1}^{T_i}y_{i,k}\log(h(k|x_i))+(1-y_{i,k})\log(1-h(k|x_i))\right\}\times\frac{1}{p(y_i=1|x_i)},
\end{equation}
\begin{equation}
\label{ylogit}
y_{i,k}= \begin{cases}
1, \quad k=T_i \\
0, \quad \text{otherwise}
\end{cases},
\end{equation}
\begin{equation}
{p(y_i=1|x_i)}=1-\sum_{t=1}^{\infty}p_t(t_i | x_i,  \theta_t)F_c(t_i|x_i,\theta_c)
\end{equation}
where $n_{(s_i=1)}$ represents the sample size of the data with $s_i=1$ and we assume that the distribution of censoring times is known in order to estimate $\theta_t$ as in the case of continuous time.
\subsubsection{Estimation Methods on Machine Learning }\label{mlest}
We introduce the following update procedure, assuming $\theta_c$ is known to estimate the model corresponding to the PU structure from the loss function proposed above. 
Loss function in Step2 and $p(y_i=1|x_i)$ in Step1 corresponds to equation(\ref{losscox}),(\ref{ycox}) respectively in the case the obeserbation is continuous, on the other hand equation (\ref{losslogit}),(\ref{ylogit}) respectively in the case the obeserbation is discrete.
\begin{table}[h]
\centering
\caption{Estimation from the Loss Function}
\small 
\begin{tabular}{|l|}
\hline
Input: dataset D \\
Output: $\hat{\theta_t}$\\
Let the initial value be $\theta_t^{(0)}.$\\
$\newline$\\
Step1:$\hspace{3mm}$ Calculate $p(y_i=1|x_i)$\\
$\newline$\\
Step2:$\hspace{3mm}$$\hat{\theta_t}$$\leftarrow$$\argmin_{\theta_t}(loss);$\\
$\newline$\\
This process is repeated until the update range of $\hat{\theta_t}$ falls below a certain level.\\
\hline
\end{tabular}
\end{table}

\section{Simulation}
In this chapter, by simulation we show that the two likelihood functions proposed in the previous chapter are correct. We first introduce the process of generating the dataset $D$ for the simulation. We then summarize the parametric methods used to estimate the model, the estimation results and the interpretation of the results.
\subsection{Data Generating Process}
First, the process of generating the dataset $D=\{\mathbf{t},\mathbf{c},\mathbf{s},\mathbf{x}\}$ is described. The following steps are used to generate the data necessary for the model. In this process, $t_i$ and $c_i$ are generated from an exponential distribution and $x_i$ is generated from a 2-dimensional multivariate normal distribution. The Python module numpy was used to generate the following data.
\begin{enumerate}
\item Determine the true values of the parameters $\theta_t,\theta_c$ for the model of survival time and the model of censoring time and the mean and covariance variance matrix of the multivariate normal distribution of $x_i$.
\item Generate $x_i$ and calculate $\lambda_{it},\lambda_{ic}$.
\item Generate $t_i,c_i$ and determine $y_i$.
\item The generated data is divided into two parts and the dataset to be labeled with $s_i=1$ is $D_1$ and the dataset to be labeled with $s_i=0$ is $D_2$.
\item We randomly label only users satisfying $y_i=1$ with $s_i=1$. from $D_1$(where we set 50\% of the dataset as $s_i=1$).Only users with $s_i = 1$ are adopted as a dataset.
\item We randomly label a part of the users with $s_i=0$. from $D_2$(where we set 50\% of the dataset as $s_i=0$).Only users with $s_i = 0$ are adopted as a dataset.
\item Combine the data taken from $D_1$ and $D_2$ to complete the final dataset D.
\end{enumerate}
A detailed description of the above operations is added. In step 1, $\theta_t,\theta_c,x_i$ is treated as a two-dimensional variable. The true values of $\theta_t,\theta_c$ are $(2,1),(1,0.5)$ respectively, the mean of the multivariate normal distribution of $x$ is (0.7,0.4), and the covariance matrix is ([0.3,-0.1],[-0.1,0.2]). In steps 2 and 3, based on the initial settings, we generated samples of $t_i,c_i,x_i$ with n=10000, and the covariance matrix was set to ([0.3,-0.1],[-0.1,0.2]). In step 3, the calculation was based on Equation (\ref{indicator}),(\ref{survival}) and (\ref{censoring}). Steps 4, 5, 6, and 7 generated a sample with $s_i$ labeling based on the two assumptions for the data structure given in Chapter 2. The labeling of $s_i$ must be done on a randomly selected sample from $y=1$, and the labeling of $s_i=0$ must be randomly selected from the whole set. For this reason, we divided the generated dataset into two parts and performed the operation shown above on each dataset $D_1,D_2$. It is clear that if $s_i$ is labeled without splitting the data set, the data set will be labeled in violation of one of the assumptions. Table 3 shows the data generated by performing this operation once. The number of samples generated was n=4126, and the number of samples labeled $s_i=1$ was n=1682 and $s_i=0$ was n=2445.
\begin{table}[h]
\centering
\caption{Generated Dataset}
\small 
\begin{tabular}{|c|c|c|c|c|c|c|c|} \hline
ID& $t_i$ & $c_i$& $x_i$&$\lambda_{it}$ & $\lambda_{ic}$ &$y_i$ &$s_i$
\\
\hline\hline
1&0.035 & 0.178 & ( 1.663, -0.045) & 26.565 & 5.154 & 1 & 1 \\
2&0.192 & 0.271 & ( 1.019, 0.393) & 11.531 & 3.316 &1 &1 \\
3&0.021& 0.090 & (1.158, 0.474) &16.557 & 3.956 &1 &1 \\
4& 0.155& 0.301 & (0.173, 1.399) & 6.097 & 2.337 & 1 &1 \\ 
\hline
$\vdots$& $\vdots$& $\vdots$&$\vdots$ &$\vdots$&$\vdots$ &$\vdots$ &$\vdots$ \\ 
\hline
4123&0.200 & 0.037 & (0.259, 0.057) &1.782 & 1.329 &0 &0 \\
4124& 0.281 & 0.764 & (-0.109, 1.088) & 2.508 & 1.520 &1 &0 \\
4125& 0.552 & 0.247 & ( 0.684, -0.223) & 3.101 & 1.766 &0 &0 \\
4126& 0.400 & 1.113 & ( 0.383, 0.489) & 3.578 & 1.850 &1 &0 \\
\hline
\end{tabular}
\end{table}
\subsection{Method}
In this section, we introduce the method and the results of estimating the true parameter $\theta_t,\theta_c$. The method adopted is based on the maximum likelihood estimation method shown in Table 4.When $\theta_t$ is estimated with maximum likelihood, $\theta_c$ is fixed as $\lambda_{ic}$. Estimating maximum likelihood of $\theta_c$ is vice versa.The minus log-likelihood of equations (\ref{t_likelihood}),(\ref{c_likelihood}) represents $-\log(L(\theta_t)|.)),-\log(L(\theta_c|.))$ respectively. Since the minus log-likelihoods were not stable when estimating $\theta_t,\theta_c$ at the same time, we used the maximum likelihood estimation alternately as shown in Table 4. The initial value is $\theta_t^{(0)}=(0,0), \theta_c^{(0)}=(0,0)$. We implemented this estimator by optimize.fmin$\_$bfgs, which is included in package scipy of Python. In the next section, we summarize the results using the estimator presented in this section.
\begin{table}[h]
\centering
\caption{Maximum Likelihood Estimator}
\small 
\begin{tabular}{|l|}
\hline
Input: dataset D \\
Output: $\hat{\theta_t},\hat{\theta_c}$\\
Put $\theta_t^{(0)}, \theta_c^{(0)}$ as Intial values\\

Step1:$\hspace{3mm}$$\hat{\theta_t}$$\leftarrow$$\argmin_{\theta_t}\left\{ -\log(L(\theta_t)|.)\right\};$\\
$\hspace{12mm}$$\lambda_{it} ^{new}$$\leftarrow$$\exp(x'\hat{\theta_t})$\\
$\newline$\\
Step2:$\hspace{3mm}$$\hat{\theta_c}$$\leftarrow$$\argmin_{\theta_c} \left\{-\log(L(\theta_c)|.)\right\};$\\
$\hspace{12mm}$$\lambda_{ic} ^{new}$$\leftarrow$$\exp(x'\hat{\theta_c})$\\
Continue until the update ranges of $\hat{\theta_t}$,$\hat{\theta_c}$ fall below a certain level.\\
\hline
\end{tabular}
\end{table}
\subsection{Result}
In this section, the estimation results of the proposed model and the conventional model are summarized in Tables 5, 6, 7 and 8 for two patterns: large (n=10000) and small (n=3000) samples. The present model is compatible with positive unlabeled data, but the conventional model is not able to handle positive unlabeled data. Therefore, the conventional model assumes the likelihood that $s_i$ is observed as a censoring indicator and the true censoring indicator $y_i$ is not known. In other words, the conventional model fails to correctly respond to the positive unlabeled data used for this comparison and recognizes $s_i$ as $y_i=0$.For comparison, we adopted the exponential model estimation method \shortciteA{lee2003statistical} for the parametric model. The likelihood functions $L_{conv}$,$L'_{conv}$ assumed in the conventional model for the case where $c$ is observed and for the case where $c$ is not observed are shown in Appendix 5,6.\par
We also used three indexes to evaluate the estimation of $\theta_{t1},\theta_{t2},\theta_{c1},\theta_{c2}$: asymptotic standard error, coverage, and RMSE. In order to obtain these indices, we created 1000 datasets and used the estimation method introduced in the previous section. When n=10000, the mean of samples was n=4250.603 and when n=3000, the mean of samples was n=1275.583. The mean of the inverse of the expected value of the observed Fisher information matrix was used to calculate the asymptotic standard error. Calculating coverage, 95\% confidence intervals were created using the observed Fisher information matrix for each dataset, and the percentage of true values in the confidence intervals was calculated, respectively. The quadratic derivative Q of each likelihood used to obtain the Fisher information matrix is summarized in Appendix 3-6. At the end of this section, the estimation accuracy of each model for the large sample is visualized in Figure 1-4, showing the boxplot at n=10000 for each of $\theta_{t1},\theta_{t2},\theta_{c1},\theta_{c2}$.

\par
In terms of RMSE, PUSA is excellent in all cases where $c$ is observed and not observed, and the RMSE ratios are also excellent in all cases of PUSA is very low.
\par
Similarly, Table 6 shows that the proposed PUSA method is much more stable than Conventional PM, which does not assume the PU structure. Although the RMSE is larger than that for $n=10,000$, the coverage rate is still about 90\%. On the other hand, Conventional PM has almost no true value.
\begin{center} 
\begin{threeparttable}[h]
\centering
\caption{Estimated results of $\theta_t,\theta_c$(n=10000)}
\small 
\begin{tabular}{|c|c|c||c|c||c|c|}
\hline
\multicolumn{3}{|l||}{} & \multicolumn{2}{c||}{PUSA\tnote{$\star$}} & \multicolumn{2}{c|}{Conventional PM\tnote{*}} \\ \hline
\multicolumn{3}{|l||}{c setting} & c observable & c unobservable & c observable & c unobservable \\ \hline \hline
\multirow{7}{*}{$\theta_t$} & \multirow{2}{*}{True Value} & $\theta_{t1}$ & \multicolumn{4}{c|}{2} \\ \cline{3-7} 
& & $\theta_{t2}$ & \multicolumn{4}{c|}{1} \\ \cline{2-7} 
                         & \multirow{2}{*}{Mean Value}            & $\theta_{t1}$ &     2.004        &    2.004        &  0.144            &   1.303                 \\ \cline{3-7} 
                          &                                & $\theta_{t2}$ &      0.999       &     0.999          &       -0.585           &      0.732       \\ \cline{2-7} 
                          & \multirow{2}{*}{Asymptotic SE\tnote{$\circ$}}            & $\theta_{t1}$ &     0.032        &    0.032        &  0.077            &   0.017                 \\ \cline{3-7} 
                          &                                & $\theta_{t2}$ &      0.055       &     0.055          &       0.101           &      0.026       \\ \cline{2-7} 
                          
                          & \multicolumn{2}{c||}{RMSE Rate\tnote{$\dagger$}}          &      0.029      &   0.096      &      \multicolumn{1}{c|} -        &      \multicolumn{1}{c|}   -               \\ \hline  \hline
\multirow{7}{*}{$\theta_c$} & \multirow{2}{*}{True Value}            & $\theta_{c1}$ & \multicolumn{4}{c|}{1}  \\ \cline{3-7} 
                          &                                & $\theta_{c2}$ & \multicolumn{4}{c|}{0.5} \\ \cline{2-7} 
                          & \multirow{2}{*}{Mean Value}            & $\theta_{c1}$ &     1.000        &  1.000             &   1.078 & -5.000       \\ \cline{3-7} 
                          &                                & $\theta_{c2}$ &    0.501         &    0.500           &    0.549      &       -6.470         \\ \cline{2-7} 
                          & \multirow{2}{*}{Asymptotic SE}            & $\theta_{c1}$ &     0.017        &  0.020             &                 0.016 & 0.084              \\ \cline{3-7} 
                          &                                & $\theta_{c2}$ &    0.025         &    0.027           &    0.0250              &       0.091      \\ \cline{2-7} 
                          
                          & \multicolumn{2}{c||}{RMSE Rate}          &      0.311     &   0.005      &     \multicolumn{1}{c|}  -           &      \multicolumn{1}{c|}   -  \\ \hline
\end{tabular}
\begin{tablenotes}
\item[$\circ$] Asymptotic SE stands for asymptotic standard error.
\item[$\star$] PUSA stands for positive Unbiased estimator for Survival Analysis.
\item[*] Conventional PM stands for Parametric estimator with exponential model.
\item[$\dagger$] RMSE Rate stands for RMSE Ratio of PUSA to Conventional PM. 
\end{tablenotes}
\end{threeparttable}
\end{center}

\begin{center} 
\begin{threeparttable}[h]
\centering
\caption{Estimated results of $\theta_t,\theta_c$(n=3000)}
\small 
\begin{tabular}{|c|c|c||c|c||c|c|}
\hline
\multicolumn{3}{|l||}{} & \multicolumn{2}{c||}{PUSA\tnote{$\star$}} & \multicolumn{2}{c|}{Conventional PM\tnote{*}} \\ \hline
\multicolumn{3}{|l||}{c setting} & c observable & c unobservable & c observable & c unobservable \\ \hline \hline
\multirow{7}{*}{$\theta_t$} & \multirow{2}{*}{True Value} & $\theta_{t1}$ & \multicolumn{4}{c|}{2} \\ \cline{3-7} 
& & $\theta_{t2}$ & \multicolumn{4}{c|}{1} \\ \cline{2-7} 
& \multirow{2}{*}{Mean Value} & $\theta_{t1}$ & 2.003 & 2.003 & 0.063 & 1.304 \\ \cline{3-7} 
& & $\theta_{t2}$ & 0.997 & 0.997 & -0.661 & 0.731 \\ \cline{2-7} 
& \multirow{2}{*}{Asymptotic SE\tnote{$\circ$}} & $\theta_{t1}$ & 0.059 & 0.059 & 0.155 & 0.031 \\ \cline{3-7} 
& & $\theta_{t2}$ & 0.101 & 0.101 & 0.196 & 0.047 \\ \cline{2-7} 
& \multicolumn{2}{c||}{RMSE Rate\tnote{$\dagger$}} & 0.050 & 0.176 & \multicolumn{1}{c|} - & \multicolumn{1}{c|} - 
\\ \hline \hline 
\multirow{7}{*}{$\theta_c$} & \multirow{2}{*}{True Value} & $\theta_{c1}$ & \multicolumn{4}{c|}{1} \\ \cline{3-7} 
& & $\theta_{c2}$ & \multicolumn{4}{c|}{0.5} \\ \cline{2-7} 
& \multirow{2}{*}{Mean Value} & $\theta_{c1}$ & 1.002 & 1.001 & 1.081 & -7.074 \\ \cline{3-7} 
& & $\theta_{c2}$ & 0.500 & 0.501 & 0.549 & -8.677 \\ \cline{2-7} 
& \multirow{2}{*}{Asymptotic SE} & $\theta_{c1}$ & 0.024 & 0.036 & 0.030 & 0.150 \\ \cline{3-7} 
& & $\theta_{c2}$ & 0.036 & 0.049 & 0.046 & 0.161 \\ \cline{2-7} 
& \multicolumn{2}{c||}{RMSE Rate} & 0.455 & 0.006 & \multicolumn{1}{c|} - & \multicolumn{1}{c|} - \\ \hline
\end{tabular}
\end{threeparttable}
\end{center}
\par
The results in Tables 7 and 8 show that the coverage rate and average coverage rate of the proposed PUSA method are clearly superior to those of Conventional PM without the PU structure, which is close to 95\% and 90\%. In particular, when c is observed with the proposed PUSA method, it is better than other methods for both large and small samples.
\begin{center} 
\begin{threeparttable}[h]
\centering
\caption{Estimated CoverageRate results with confidence intervals of 95\% and 90\ of $\theta_t,\theta_c$(n=10000)}
\small 
\begin{tabular}{|c|c|c||c|c||c|c|}
\hline
\multicolumn{3}{|l||}{} & \multicolumn{2}{c||}{PUSA\tnote{$\star$}} & \multicolumn{2}{c|}{Conventional PM\tnote{*}} \\ \hline
\multicolumn{3}{|l||}{c setting} & c observable & c unobservable & c observable & c unobservable \\ \hline \hline
\multirow{4}{*}{95\% Coverage Rate\tnote{$\diamond$}} & \multirow{2}{*}{$\theta_t$} & $\theta_{t1}$ &     0.920        &      0.917         &                 0.0 &      0.0              \\ \cline{3-7} 
                          &                                & $\theta_{t2}$ &    0.931         &     0.929          &     0.0             &    0.0                 \\ \cline{2-7} 
& \multirow{2}{*}{$\theta_c$} & $\theta_{c1}$ &   0.940          &    0.893           &   0.078               &     0.0               \\ \cline{3-7} 
                          &                                & $\theta_{c2}$ &   0.935          &     0.875          &     0.509             &       0.0             \\ \hline
                          \multicolumn{3}{|c||}{95\% Average Coverage Rate\tnote{$\ddagger$}}          &      0.9315    &  0.9035       &    0.14675 &      0.0      \\ \hline \hline
                          
\multirow{4}{*}{90\% Coverage Rate\tnote{$\diamond$}} & \multirow{2}{*}{$\theta_t$} & $\theta_{t1}$ &     0.854        &      0.852         &                 0.0 &      0.0              \\ \cline{3-7} 
                          &                                & $\theta_{t2}$ &    0.859         &     0.862          &     0.0             &    0.0                \\ \cline{2-7}
& \multirow{2}{*}{$\theta_c$} & $\theta_{c1}$ &   0.882          &    0.823           &   0.057              &     0.0               \\ \cline{3-7} 
                          &                                & $\theta_{c2}$ &   0.858          &     0.787          &     0.423             &       0.0             \\ \hline
                          \multicolumn{3}{|c||}{90\%Average Coverage Rate\tnote{$\ddagger$}}          &   0.86325 &  0.831       &     0.12         &     0.0       \\ \hline
\end{tabular}
\begin{tablenotes}
\item[$\diamond$] 95\% Coverage Rate stands for Percentage of the true value in the 95\% confidence interval.
\item[$\ddagger$]95\% Average Coverage Rate stands for the average of the four 95\% coverage rate.
\end{tablenotes}
\end{threeparttable}
\end{center}

\begin{center} 
\begin{threeparttable}[h]
\centering
\caption{Estimated CoverageRate results with confidence intervals of 95\% and 90\ of $\theta_t,\theta_c$(n=3000)}
\small 
\begin{tabular}{|c|c|c||c|c||c|c|}
\hline
\multicolumn{3}{|l||}{} & \multicolumn{2}{c||}{PUSA\tnote{$\star$}} & \multicolumn{2}{c|}{Conventional PM\tnote{*}} \\ \hline
\multicolumn{3}{|l||}{c setting} & c observable & c unobservable & c observable & c unobservable \\ \hline \hline
\multirow{4}{*}{95\% Coverage Rate\tnote{$\diamond$}} & \multirow{2}{*}{$\theta_t$} & $\theta_{t1}$ & 0.918 & 0.919 & 0.0 & 0.0 \\ \cline{3-7} 
& & $\theta_{t2}$ & 0.907 & 0.899 & 0.0 & 0.001 \\ \cline{2-7} 
& \multirow{2}{*}{$\theta_c$} & $\theta_{c1}$ & 0.928 & 0.888 & 0.376
& 0.0 \\ \cline{3-7} 
& & $\theta_{c2}$ & 0.927 & 0.881 & 0.689 & 0.0 \\ \hline
\multicolumn{3}{|c||}{95\%Average Coverage Rate\tnote{$\ddagger$}} & 0.92 & 0.897 & 0.26625 & 0.00025 \\ \hline \hline
\multirow{4}{*}{90\% Coverage Rate\tnote{$\diamond$}} & \multirow{2}{*}{$\theta_t$} & $\theta_{t1}$ & 0.854 & 0.842 & 0.0 & 0.0 \\ \cline{3-7} 
& & $\theta_{t2}$ & 0.844 & 0.832 & 0.0 & 0.0 \\ \cline{2-7} 
& \multirow{2}{*}{$\theta_c$} & $\theta_{c1}$ & 0.874 & 0.818 & 0.316
& 0.0 \\ \cline{3-7} 
& & $\theta_{c2}$ & 0.867 & 0.808 & 0.603 & 0.0 \\ \hline
\multicolumn{3}{|c||}{90\% Average Coverage Rate\tnote{$\ddagger$}} & 0.860 & 0.825 & 0.230 & 0.0 \\ \hline
\end{tabular}
\end{threeparttable}
\end{center}
\par
The boxplots for each model at n=10000 for $\theta_{t1},\theta_{t2},\theta_{c1},\theta_{c2}$ are summarized in Figure 1-4. Each x-axis shows, from left to right, the cases where c is or is not observed in PUSA and the cases where c is or is not observed in the Conventional PM, respectively.
\begin{figure}
\begin{tabular}{cc}
\begin{minipage}{0.5\hsize}
\begin{center}
\includegraphics[scale=0.4]{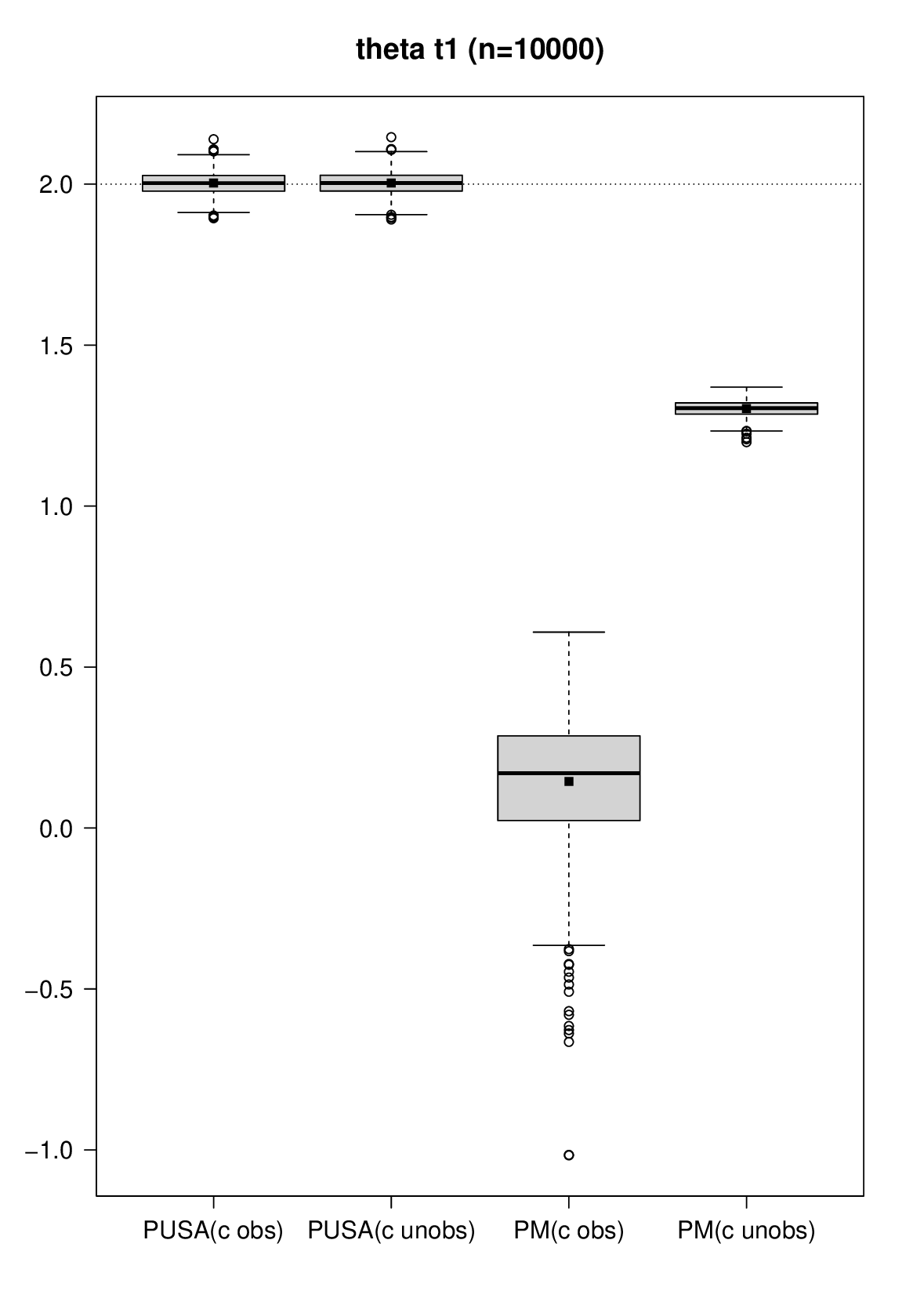}
\caption{Comparison of the estimation accuracy of $\theta_{t1}$}
\end{center}
\end{minipage}
\begin{minipage}{0.5\hsize}
\begin{center}
\includegraphics[scale=0.4]{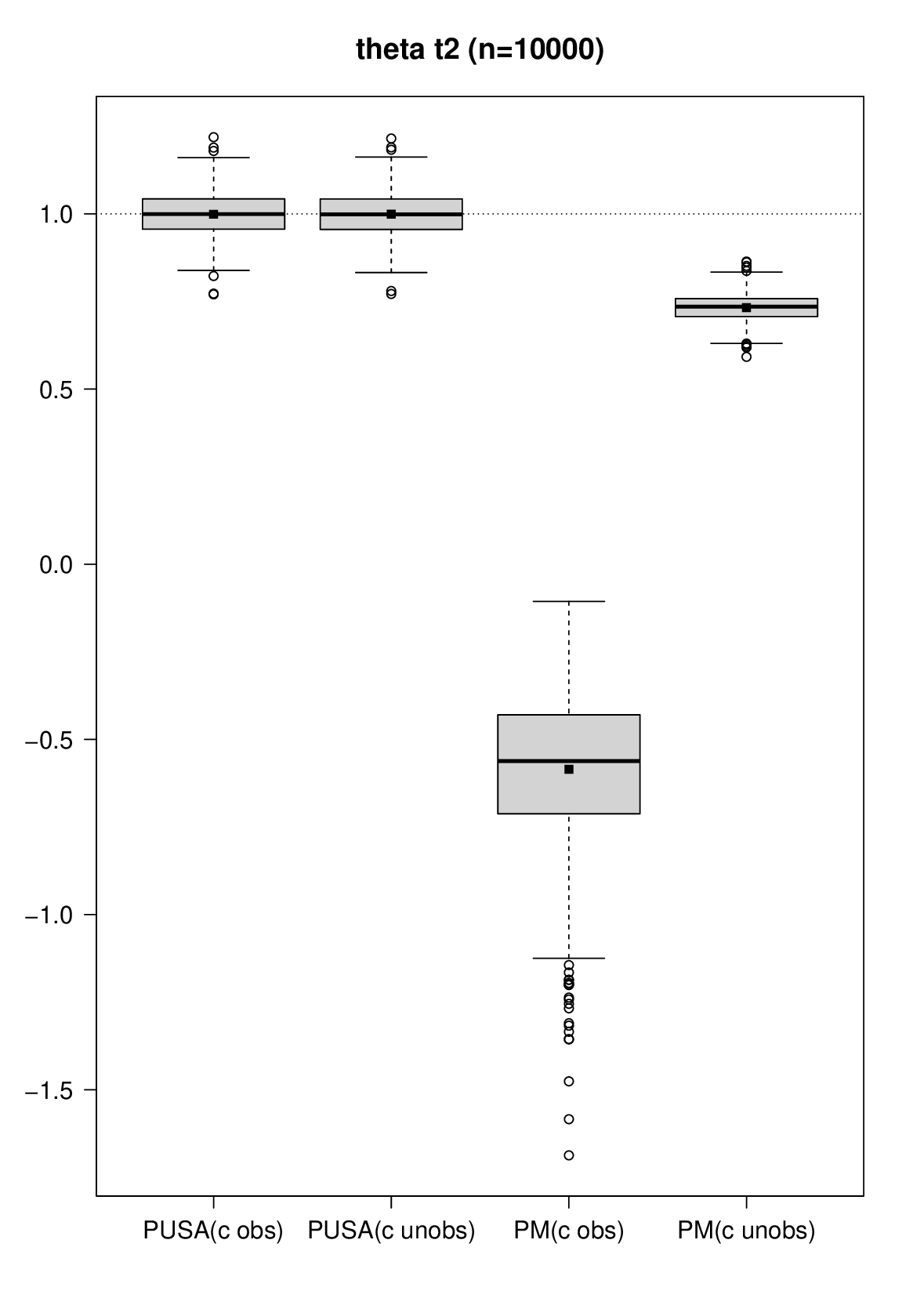}
\caption{Comparison of the estimation accuracy of $\theta_{t2}$}
\end{center}
\end{minipage}
\end{tabular}
\begin{tabular}{cc}
\begin{minipage}{0.5\hsize}
\begin{center}
\includegraphics[scale=0.4]{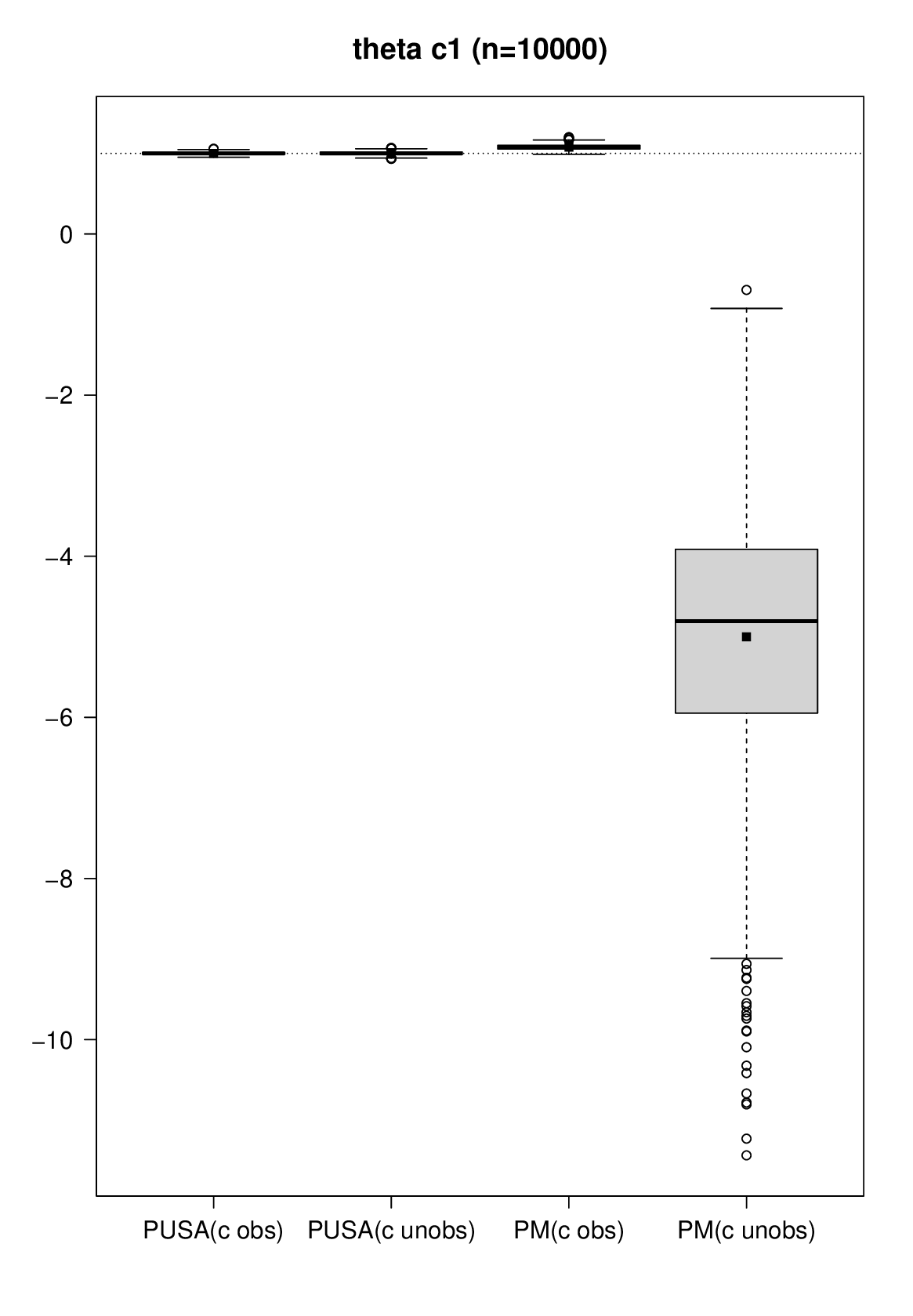}
\caption{Comparison of the estimation accuracy of $\theta_{c1}$}
\end{center}
\end{minipage}
\begin{minipage}{0.5\hsize}
\begin{center}
\includegraphics[scale=0.4]{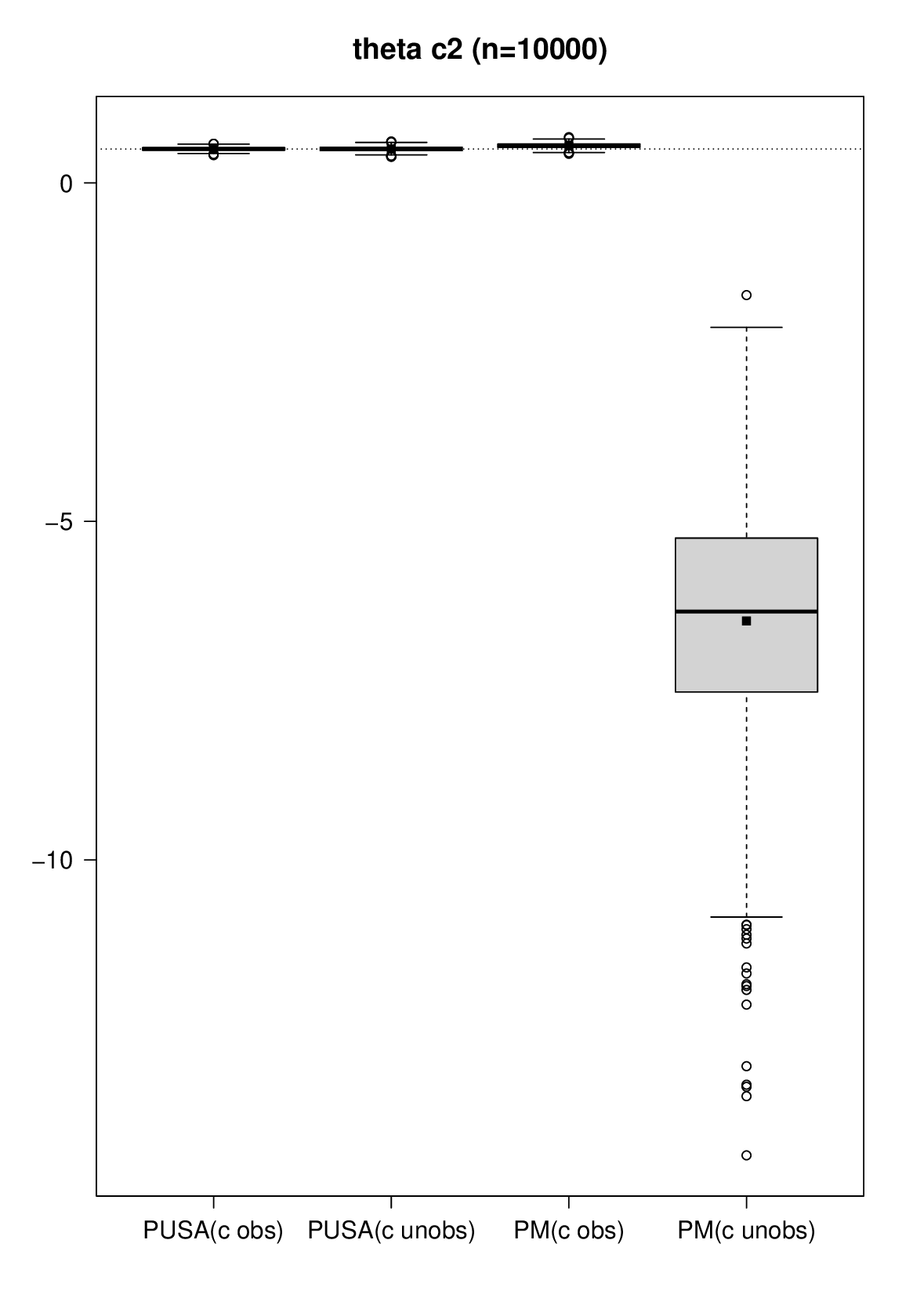}
\caption{Comparison of the estimation accuracy of $\theta_{c2}$}
\end{center}
\end{minipage}
\end{tabular}
\end{figure}

\section{Conclusion}
In this section, we summarize the results of the simulations performed in the previous section. It is clear from Table 5 that the model proposed here for the data containing $s_i$, which is shown to be uncertain, results in results that are closer to the true value than the existing models. The estimated value of $\theta_c$ in the Conventional PM when all $c$ is observed is close to the true value, but it is not even higher than that of PUSA when $c$ is partially absent in the mean square error.
This is due to the fact that the likelihood function assumed in our model is for positive unlabeled data.

As a future prospect, we proposed a method based on maximum likelihood estimation for the likelihood function, but it is expected to be used to estimate the model by the MCMC method, which is a Bayesian estimation method. Although we have assumed exponential distributions for the models of survival time and censoring time, it is expected that we can estimate more complicated models such as Weibull distribution by integrating numerically.

\section*{Appendix}
\subsection*{Appendix1}
In the case of exponential distribution, the results of the integration that appeared in this paper are summarized below.
\begin{align*}
&1-\int_0^{\infty} p_t(t_i|x_i,\theta_t)F_c(t_i|x_i,\theta_c)dt_i\\
=&1-\int_{0}^{\infty} f(t_i|x_i,\theta_t) \int_{0}^{t_i}g(c_i|x_i,\theta_c)dc_i dt_i\\
=&1-\int_{0}^{\infty} \lambda_{it} \exp(-\lambda_{it} t_i) \int_{0}^{t_i}\lambda_c \exp(-\lambda_{ic} c_i)dc_i dt_i\\
=&1-\int_{0}^{\infty} \lambda_{it} \exp(-\lambda_{it} t_i) (1-\exp(-\lambda_{ic} t_i)) dt_i\\
=&\frac{\lambda_{it}}{\lambda_{it}+\lambda_{ic} }
\end{align*}
\begin{align*}
& p_t(t_i|x_i,\theta_t) \{1-F_c(t_i|x_i,\theta_c)\}\\
=& \lambda_{it} \exp(-\lambda_{it} t_i )\left\{1-\int_{0}^{t_i} \lambda_{ic} \exp( \lambda_{ic} c_i)dc_i\right\}\\
=& \lambda_{it} \exp(-t_i(\lambda_{it}+\lambda_{ic}))
\end{align*}

\subsection*{Appendix2}
\subsubsection*{A case where both are gamma distribution}
The regression model of the probability density functions of survival time t and censoring time c, as the gamma distribution of parameters $\lambda_{it},\lambda_{ic},\alpha_t,\alpha_c$, respectively, with the parameters depending on x is as follows.
\[p_t(t_i | x_i, \alpha_t, \theta_t) = \frac{1}{\Gamma(\alpha_t)}t_i^{\alpha_t-1}\lambda_{it}^{\alpha_t } \exp( -\lambda_{it} t_i), \quad \lambda_{it} = \exp(x_i' \theta_t) \]
\[p_c(c_i | x_i, \alpha_c, \theta_c) = \frac{1}{\Gamma(\alpha_c)}c_i^{\alpha_c-1}\lambda_{ic}^{\alpha_c} \exp( -\lambda_{ic} c_i), \quad \lambda_{ic} = \exp(x_i' \theta_c) \]

\[
F_c(c_i | x_i, \alpha_c, \theta_c) = \frac{\gamma(\alpha_c, \lambda_{ic}c_i)}{\Gamma(\alpha_c)}
\]
where $\gamma(a,b)$ stands for (first-type) Incomplete gamma function,then
\begin{align*}
&1-\int_{0}^{\infty} p_t(t_i|x_i,\theta_t)F_c(t_i|x_i,\theta_c)dt_i\\
=&1-\int_{0}^{\infty} \frac{1}{\Gamma(\alpha_t)}t_i^{\alpha_t-1}\lambda_{it}^{\alpha_t } \exp( -\lambda_{it} t_i) \frac{\gamma(\alpha_c, \lambda_{ic}c_i)}{\Gamma(\alpha_c)} dt_i
\end{align*}
However, this is analytically unintegrable.
\subsubsection*{A case where both are weibull distribution}
The regression model of the probability density functions of survival time t and censoring time c, as the weibull distribution with parameters $\lambda_{it},\lambda_{ic},\alpha_t,\alpha_c$, respectively, with the parameters depending on x is as follows.
\[p_t(t_i | x_i, \alpha_t, \theta_t) = \alpha_t t_i^{\alpha_t-1} \lambda_{it} \exp(-\lambda_{it} t_i^{\alpha_t}), \quad \lambda_{it} = \exp(x_i \theta_t) \]
\[p_c(c_i | x_i, \alpha_c, \theta_c) = \alpha_c c_i^{\alpha_c-1} \lambda_{ic} \exp(-\lambda_{ic} c_i^{\alpha_c}), \quad \lambda_{ic} = \exp(x_i \theta_c) \]

\[
F_c(c_i | x_i, \alpha_c, \theta_c) = 1 - \exp(-\lambda_{ic}c_i^{\alpha_c})
\]

Therefore,
\begin{align*}
&1-\int_0^{\infty} p_t(t_i | x_i, \alpha_t, \theta_t)F_c(t_i|x_i, \alpha_c, \theta_c)dt\\
= &1-\int_0^{\infty} \alpha_t t_i^{\alpha_t-1} \lambda_{it} \exp(-\lambda_{it} t_i^{\alpha_t})(1 - \exp(-\lambda_{ic}t_i^{\alpha_c}))dt
\end{align*}
If not $\alpha_t=\alpha_c$, then it can not be integrable

\subsubsection*{A case where Weibull distribution and exponential distribution respectively}
A regression model in which the parameter depends on $x$ with the probability density function of survival time t as the Weibull distribution with parameters $\lambda_{it},\alpha_t$ and the probability density function of censoring time c as the exponential distribution with parameter $\lambda_{ic}$ is shown below.
\[p_t(t_i | x_i, \alpha_t, \theta_t) = \alpha_t t_i^{\alpha_t-1} \lambda_{it} \exp(-\lambda_{it} t_i^{\alpha_t}), \quad \lambda_{it} = \exp(x_i \theta_t) \]
\[p_c(c_i | x_i, \theta_c) = \lambda_{ic} \exp(-\lambda_{ic} c_i), \quad \lambda_{ic} = \exp(x_i' \theta_c)\]

\[
F_c(c_i | x_i, \alpha_c, \theta_c) = 1 - \exp(-\lambda_{ic}c_i)
\]

Therefore,
\begin{align*}
&1-\int_0^{\infty} p_t(t_i | x_i, \alpha_t, \theta_t)F_c(t_i|x_i, \alpha_c, \theta_c)dt\\
= &1-\int_0^{\infty} \alpha_t t_i^{\alpha_t-1} \lambda_{it} \exp(-\lambda_{it} t_i^{\alpha_t})(1 - \exp(-\lambda_{ic}t_i))dt
\end{align*}
If not $\alpha_t=\alpha_c$, then it can not be integrable

\subsection*{Appendix3}
The likelihood and log-likelihood functions when $c$ is observed are as follows.
\begin{align*}
L(\theta_t|.) & \propto \prod^n_{i=1} \{(\lambda_t+\lambda_c) \exp(-\lambda_t t_i)\}^{s_i}\\
\log L(\theta_t|.) &= \sum_{i=1}^n s_i (\log(\lambda_{it}+\lambda_{ic}) -\lambda_{it}t_i)\\
L(\theta_c|.) & \propto \prod^n_{i=1} (\lambda_{it}+\lambda_{ic})^{s_i}\lambda_{ic} \exp(-\lambda_{ic} c_i)\\
\log L(\theta_c | .) &= \sum_{i=1}^n s_i \log(\lambda_{it}+\lambda_{ic}) + x_i' \theta_c-\lambda_{ic} c_i
\end{align*}

The partial derivatives are as follows.
\begin{align*}
g(\theta_t)=\nabla_{\theta_t} \log L(\theta_t | .) &=\sum_{i=1}^n s_i x_i \left(\frac{ \lambda_{it}}{\lambda_{it}+\lambda_{ic}}-\lambda_{it}t_i\right)\\
Q(\theta_t)=- \nabla_{\theta_t} \nabla_{\theta_t}' \log L(\theta_t | .) &=\sum_{i=1}^n -s_ix_i x_i'\left\{\frac{\lambda_{it}}{\lambda_{it}+\lambda_{ic}}\left(1-\frac{\lambda_{it}}{\lambda_{it}+\lambda_{ic}}\right)-\lambda_{it}t_i\right\}\\
g(\theta_c)=\nabla_{\theta_c} \log L(\theta_c | .) &=\sum_{i=1}^n x_i\left(s_i\frac{\lambda_{ic}}{\lambda_{it}+\lambda_{ic}}+1-\lambda_{ic}c_i\right)\\
Q(\theta_c)=- \nabla_{\theta_c} \nabla_{\theta_c}' \log L(\theta_c | .) &=\sum_{i=1}^n -x_i x_i'\left\{s_i\frac{\lambda_{ic}}{\lambda_{it}+\lambda_{ic}}\left(1-\frac{\lambda_{ic}}{\lambda_{it}+\lambda_{ic}}\right)-\lambda_{ic}c_i\right\}\\
\end{align*} 

\subsection*{Appendix4}
The likelihood and log-likelihood functions when $c$ is unobserved are as follows.
\begin{align*}
L(\theta_t|.) & \propto \prod^n_{i=1} \left\{(\lambda_{it}+\lambda_{ic})\exp(-\lambda_{it}t_i)\right\}^{s_i}\\
\log L(\theta_t|.) &= \sum_{i=1}^n s_i (\log(\lambda_{it}+\lambda_{ic}) -\lambda_{it}t_i)\\
L(\theta_c|.) & \propto \prod^n_{i=1} \{(\lambda_{it}+\lambda_{ic})\exp(-\lambda_{ic}t_i)\}^{s_i}\{\lambda_{ic} \exp(-\lambda_{ic} c_i)\}^{1-s_i}\\
\log L(\theta_c | .) &= \sum_{i=1}^n s_i (\log(\lambda_{it}+\lambda_{ic}) -\lambda_{ic}t_i) + (1-s_i)(x_i' \theta_c-\lambda_{ic}c_i)
\end{align*} 

The partial derivatives are as follows.
\begin{align*}
g(\theta_t)=\nabla_{\theta_t} \log L(\theta_t | .) &=\sum_{i=1}^n s_i x_i \left(\frac{ \lambda_{it}}{\lambda_{it}+\lambda_{ic}}-\lambda_{it}t_i\right)\\
Q(\theta_t)=- \nabla_{\theta_t} \nabla_{\theta_t}' \log L(\theta_t | .) &=\sum_{i=1}^n -s_ix_i x_i'\left\{\frac{\lambda_{it}}{\lambda_{it}+\lambda_{ic}}\left(1-\frac{\lambda_{it}}{\lambda_{it}+\lambda_{ic}}\right)-\lambda_{it}t_i\right\}\\
g(\theta_c)=\nabla_{\theta_c} \log L(\theta_c | .) &=\sum_{i=1}^n x_i \left\{s_i\left(\frac{\lambda_{ic}}{\lambda_{it}+\lambda_{ic}}-\lambda_{ic}t_i\right)+(1-s_i)(1-\lambda_{ic}c_i)\right\}\\
Q(\theta_c)=- \nabla_{\theta_c} \nabla_{\theta_c}' \log L(\theta_c | .) &=\sum_{i=1}^n -x_i x_i' \left[s_i\left\{\frac{\lambda_{ic}}{\lambda_{it}+\lambda_{ic}}\left(1-\frac{\lambda_{ic}}{\lambda_{it}+\lambda_{ic}}\right)-\lambda_{ic}t_i\right\}-(1-s_i)\lambda_{ic}c_i\right]\\
\end{align*}

\subsection*{Appendix5}
When $c$ is observed, the likelihood function without considering the PU structure and the log-likelihood function is as follows.
\begin{align*}
&\prod_{i=1}^n p(t_i, c_i | s_i=1, x_i)^{s_i}p(c_i | s_i = 0, x_i)^{1-s_i}\\
=& \prod_{i=1}^n \left\{\frac{p_t(t_i|x_i,\theta_t)p_c(c_i|x_i,\theta_c)}{1-\int_0^{\infty} p_t(t_i|x_i,\theta_t)F_c(t_i|x_i,\theta_c)dt_i}\right\}^{s_i}\left\{\frac{1-p_t(t_i|x_i,\theta_t) [1-F_c(t_i|x_i,\theta_c)]}{\int_0^{\infty} p_t(t_i|x_i,\theta_t)F_c(t_i|x_i,\theta_c)dt_i}\right\}^{1-s_i}\\
=& \prod_{i=1}^n \left\{\lambda_{ic} \exp(-\lambda_{ic} c_i)\exp(-\lambda_{it} t_i)\right\}^{s_i}\{\exp(-c_i(\lambda_{it}+\lambda_{ic}))\}^{1-s_i}(\lambda_{it}+\lambda_{ic})
\end{align*}

\begin{align*}
L(\theta_t|.) &\propto \prod^n_{i=1} \exp(-\lambda_{it} t_i)^{s_i}\exp(-\lambda_{it}c_i)^{1-s_i}(\lambda_{it}+\lambda_{ic})\\
\log L(\theta_t|.) &= \sum_{i=1}^n -s_i \lambda_{it}t_i-(1-s_i) \lambda_{it}c_i+\log(\lambda_{it}+\lambda_{ic})\\
L(\theta_c|.) &\propto \prod_{i=1}^n \lambda_{ic} ^{s_i}\exp(-\lambda_{ic}c_i)(\lambda_{it}+\lambda_{ic})\\
\log L(\theta_c | .) &= \sum_{i=1}^n s_i x_i'\theta_c -\lambda_{ic}c_i +\log(\lambda_{it}+\lambda_{ic})
\end{align*} 
The partial derivatives are as follows.
\begin{align*}
g(\theta_t)=\nabla_{\theta_t} \log L(\theta_t | .) &=\sum_{i=1}^n x_i\left\{-s_i \lambda_{it}t_i-(1-s_i)\lambda_{it}c_i+\frac{ \lambda_{it}}{\lambda_{it}+\lambda_{ic}}\right\}\\
Q(\theta_t)=- \nabla_{\theta_t} \nabla_{\theta_t}' \log L(\theta_t | .) &=\sum_{i=1}^n x_i x_i'\left\{s_i \lambda_{it}t_i+(1-s_i)\lambda_{it}c_i-\frac{\lambda_{it}}{\lambda_{it}+\lambda_{ic}}\left(1-\frac{\lambda_{it}}{\lambda_{it}+\lambda_{ic}}\right)\right\}\\
g(\theta_c)=\nabla_{\theta_c} \log L(\theta_c | .) &= \sum_{i=1}^n x_i\left(s_i -\lambda_{ic}c_i +\frac{ \lambda_{ic}}{\lambda_{ic}+\lambda_{ic}}\right)\\
Q(\theta_c)=- \nabla_{\theta_c} \nabla_{\theta_c}' \log L(\theta_c | .) &=\sum_{i=1}^n x_i x_i' \left\{\lambda_{ic}c_i-\frac{\lambda_{ic}}{\lambda_{it}+\lambda_{ic}}\left(1-\frac{\lambda_{ic}}{\lambda_{it}+\lambda_{ic}}\right)\right\}\\
\end{align*} 

\subsection*{Appendix6}
When $c$ is unobserved, the likelihood function without considering the PU structure and the log-likelihood function is as follows.
\begin{align*}
&\prod_{i=1}^n p(t_i | s_i=1, x_i)^{s_i}p(c_i | s_i = 0, x_i)^{1-s_i}\\
=& \prod_{i=1}^n \left\{\frac{p_t(t_i|x_i,\theta_t) [1-F_c(t_i|x_i,\theta_c)]}{1-\int_0^{\infty} p_t(t_i|x_i,\theta_t)F_c(t_i|x_i,\theta_c)dt_i}\right\}^{s_i}\left\{\frac{1-p_t(t_i|x_i,\theta_t) [1-F_c(t_i|x_i,\theta_c)]}{\int_0^{\infty} p_t(t_i|x_i,\theta_t)F_c(t_i|x_i,\theta_c)dt_i}\right\}^{1-s_i}\\
=& \prod_{i=1}^n \exp(-t_i(\lambda_{it}+\lambda_{ic}))^{s_i}\exp(-c_i(\lambda_{it}+\lambda_{ic}))^{1-s_i}(\lambda_{it}+\lambda_{ic})
\end{align*}

\begin{align*}
L(\theta_t|.) &\propto \prod^n_{i=1} \exp(-\lambda_{it}t_i)^{s_i}\exp(-\lambda_{it}c_i)^{1-s_i}(\lambda_{it}+\lambda_{ic})\\
\log L(\theta_t|.) &= \sum_{i=1}^n -s_i \lambda_{it}t_i-(1-s_i) \lambda_{it}c_i+\log(\lambda_{it}+\lambda_{ic})\\
L(\theta_c|.) &\propto \prod^n_{i=1} \exp(-\lambda_{ic}t_i)^{s_i}\exp(-\lambda_{ic}c_i)^{1-s_i}(\lambda_{it}+\lambda_{ic})\\
\log L(\theta_c | .) &= \sum_{i=1}^n -s_i \lambda_{ic}t_i-(1-s_i) \lambda_{ic}c_i+\log(\lambda_{it}+\lambda_{ic})
\end{align*} 
The partial derivatives are as follows.
\begin{align*}
g(\theta_t)=\nabla_{\theta_t} \log L(\theta_t | .) &=\sum_{i=1}^n x_i\left\{-s_i \lambda_{it}t_i-(1-s_i)\lambda_{it}c_i+\frac{ \lambda_{it}}{\lambda_{it}+\lambda_{ic}}\right\}\\
Q(\theta_t)=- \nabla_{\theta_t} \nabla_{\theta_t}' \log L(\theta_t | .) &=\sum_{i=1}^n x_i x_i'\left\{s_i \lambda_{it}t_i+(1-s_i)\lambda_{it}c_i-\frac{\lambda_{it}}{\lambda_{it}+\lambda_{ic}}\left(1-\frac{\lambda_{it}}{\lambda_{it}+\lambda_{ic}}\right)\right\}\\
g(\theta_c)=\nabla_{\theta_c} \log L(\theta_c | .) &=\sum_{i=1}^n x_i\left\{-s_i \lambda_{ic}t_i-(1-s_i)\lambda_{ic}c_i+\frac{ \lambda_{ic}}{\lambda_{it}+\lambda_{ic}}\right\}\\
Q(\theta_c)=- \nabla_{\theta_c} \nabla_{\theta_c}' \log L(\theta_c | .) &=\sum_{i=1}^n x_i x_i'\left\{s_i \lambda_{ic}t_i+(1-s_i)\lambda_{ic}c_i-\frac{\lambda_{ic}}{\lambda_{it}+\lambda_{ic}}\left(1-\frac{\lambda_{ic}}{\lambda_{it}+\lambda_{ic}}\right)\right\}\\
\end{align*} 

\bibliography{1126_Eng_PUSA_revised}
\end{document}